\documentclass[aps, prl, twocolumn, preprintnumbers, letterpaper, 10pt]{revtex4-1}

\usepackage{amsmath, amsfonts, amssymb}
\usepackage{bm, bbm}

\newcommand{\beq}{\begin{equation}\begin{aligned}{}}
\newcommand{\eeq}{\end{aligned}\end{equation}}

\newcommand{\eql}[1]{\label{eq:#1}}
\newcommand{\eq}[1]{(\ref{eq:#1})}

\newcommand{\fr}[2]{\frac{#1}{#2}}

\newcommand{\del}{\partial}
\newcommand{\dd}{\mathrm{d}}

\newcommand{\I}{\mathrm{i}}
\newcommand{\e}{\mathrm{e}}

\newcommand{\dt}{\!\cdot\!}

\newcommand{\dsp}[1]{\displaystyle{#1}} 

\newcommand{\al}{\alpha}
\newcommand{\be}{\beta}

\newcommand{\de}{\delta}
\newcommand{\De}{\Delta}

\newcommand{\la}{\lambda}
\newcommand{\La}{\Lambda}
\newcommand{\sgm}{\sigma}

\newcommand{\cD}{\mathcal{D}}
\newcommand{\cF}{\mathcal{F}}
\newcommand{\cJ}{\mathcal{J}}
\newcommand{\cL}{\mathcal{L}}
\newcommand{\cO}{\mathcal{O}}

\newcommand{\PD}{{\phantom\dag}}

\begin{document}

\title{Nonlinearly realized conformal invariance in scale invariant field theories}

\author{Takemichi Okui}
\affiliation{Department of Physics, Florida State University, Tallahassee, FL 32306, USA,}
\email{tokui@fsu.edu}
\affiliation{Theory Center, High Energy Accelerator Research Organization (KEK), Tsukuba 305-0801, Japan}

\preprint{KEK-TH-2092}

\begin{abstract}
Implications are explored of promoting non-conformal scale-invariant theories to conformal theories by nonlinearly realizing the missing symmetry. Properties of the associated Nambu-Goldstone mode imply that conformal invariance cannot be spontaneously broken to scale invariance in unitary theories and that, as well known, scale invariant unitary theories in two dimensions are also conformal. The promoted theories have only conformal primaries and no descendants. The (non-)decoupling of the Nambu-Goldstone mode is explicitly shown in examples of scale invariant theories that are actually (not) conformal.
\end{abstract}

\maketitle

A long-standing fundamental question in conformal field theory is whether conformal invariance follows from mere scale invariance (along with some general mundane assumptions such as unitarity and Poincar\'e invariance).
While the answer is rigorously proven~\cite{Polchinski:1987dy} to be positive in two dimensions based on Zamolodchikov's $c$-theorem~\cite{Zamolodchikov:1986gt},
a rigorous non-perturbative proof is yet to be found in higher dimensions (see, e.g.,~\cite{Luty:2012ww, Dymarsky:2013pqa, Dymarsky:2014zja, Yonekura:2014tha, Naseh:2016maw} for recent developments and~\cite{Nakayama:2013is} for a review). 

I would like to study a related, but different and perhaps more modest, problem. 
I will assume that a scale invariant field theory exists that is not conformally invariant,
and explore promoting it to a conformally invariant theory by introducing a Nambu-Goldstone (NG) mode and nonlinearly realizing the action of the coset $\{\text{Conformal}\} / \{\text{Scale}\}$, \`a la the Callan-Coleman-Wess-Zumino formalism~\cite{Coleman:1969sm, *Callan:1969sn} suitably adapted to the breaking of a spacetime symmetry~\cite{Low:2001bw}. 
(Poincar\'e invariance will be assumed throughout so this should be distinguished from the ``pseudo-conformal'' scenario~\cite{Hinterbichler:2011qk, *Hinterbichler:2012mv}.)

I will point out that, in three spacetime dimensions and higher, the promoted theories are manifestly not unitary
if the conformal-to-scale breaking is spontaneous.
Conformal invariance, therefore, cannot be spontaneously broken to scale invariance in unitary theories.
In two dimensions, unitarity will manifestly force the NG mode to decouple, confirming that unbroken scale invariance implies unbroken conformal invariance.
The promoted theories will be peculiar conformal theories that contain only primary operators and no descendants.
I will also apply the present formalism to explicit examples of scale invariant theories to show that the NG mode decouples in the zero momentum limit in the theories that are also conformal,  
while it does not in the theories that are not. 

I hope that some incarnations of the properties of the NG mode and peculiarities of the promoted theories might turn out to be key elements of an eventual proof that scale invariance implies conformal invariance.
Until such a proof is provided,
the present formalism may also be useful for phenomenological studies of non-conformal scale invariance, assuming that such a theory exists. 

We begin by an elementary analysis of the coordinates of the $\{\text{Conformal}\} / \{\text{Scale}\}$ coset space. 
This part is purely group theoretical.
The only relevant aspect of field theories is that the action of $P_\mu$ is nonlinearly realized as a shift in $x^\mu$ as if Poincar\'e were broken to Lorentz.
Thus, for the breaking of conformal invariance $G$ to scale invariance $H$, 
we take $G = \{ P_\mu, J_{\mu\nu}, D, K_\mu \}$ and $H = \{ J_{\mu\nu}, D \}$.
Then, focussing on the elements of $G$ continuously connected to the identity,
we parametrize the elements of the coset, $G / H = \{ P_\mu, K_\mu \}$, as
\beq
\Omega(x, \xi) = \e^{\I x \cdot P} \e^{\I \xi \cdot K}
\eql{def:Omega} 
\eeq
with parameters $x^\mu$ and $\xi^\mu$.
(The parameter $\xi^\mu$ will be promoted to a vector field later, but not yet.)
We take the coset to be a right coset, i.e., 
for any given $g \in G$ we have
$\dsp{g \Omega(x, \xi)} = \dsp{\Omega(x', \xi') \, h}$ with some $h \in H$, or, 
\beq
\Omega(x', \xi') = g \, \Omega(x, \xi) \, h^{-1}(g, x, \xi) 
\,.\eql{trans:Omega}
\eeq
The different coset over $\{J_{\mu\nu}\}$, of course, was studied a long ago~\cite{Salam:1970qk} for the breaking of the conformal group all the way down to Poincar\'e.
The present coset~\eq{def:Omega} has also been studied in~\cite{Ivanov:1981wn, Ivanov:1981wm, Kharuk:2017jwe} for unbroken conformal group 
as that is
as useful as studying the coset~$\dsp{\e^{\I p \cdot x}}$ for unbroken Poincar\'e.
Below, we quickly review the construction of this coset before we use it to explore the breaking of the conformal to scale invariance.

We want to determine $x'$, $\xi'$, and $h$ in~\eq{trans:Omega} for $g = \dsp{\e^{\I b \cdot K}}$ with a parameter $b^\mu$. 
For such $g$, \eq{trans:Omega} becomes
\beq
g \, \Omega(x, \xi) \, h^{-1}
&= \e^{\I b \cdot K} \e^{\I x \cdot P} \e^{\I \xi \cdot K} \, h^{-1}
\\
&= \e^{\I b \cdot K} \e^{\I x \cdot P} \e^{-\I b \cdot K} \, \e^{\I (\xi+b) \cdot K} \, h^{-1}
\,,\eql{KPK0}
\eeq
because $[K_\mu, K_\nu] = 0$.
Below we will always take $b^\mu$ to be infinitesimal and ignore $\cO(b^2)$.
No generality is lost
as we are only considering the group elements continuously connected to the identity.
We can then rewrite the first three exponentials in the last line above as
\beq
\e^{\I b \cdot K} \e^{\I x \cdot P} \e^{-\I b \cdot K}
= \e^{\I x \cdot P + [\I b \cdot K,\, \I x \cdot P]}
\,.\eql{KPK1}
\eeq
To proceed, we need to establish our sign and normalization conventions for the generators.
For $D$, we fix them through the commutators:
\beq
[D, P_\mu] = -\I P_\mu
\,,\quad
[D, K_\mu] = \I K_\mu
\,,\eql{scaling:PK}
\eeq
while Lorentz invariance implies $[D, J_{\al\be}] = 0$.
We adopt the $-$$+$$+\cdots$ metric signature so the Lorentz generators for the vector representation are given by
\beq
\bigl[ \cJ_{\al\be} \bigr]^\mu_{~\nu} 
= -\I ( \de_\al^ \mu \eta_{\be\nu}^\PD - \de_\be^ \mu \eta_{\al\nu}^\PD )
\,.\eql{J:4-vec}
\eeq
The remaining commutators of $J_{\al\be}$ are then given by
\beq
&[J_{\al\be}, Z_\mu] = Z_\nu \bigl[ \cJ_{\al\be} \bigr]^\nu_{~\mu} 
\,,\\
&[J_{\al\be}, J_{\mu\nu}] 
= J_{\rho\nu} \bigl[ \cJ_{\al\be} \bigr]^\rho_{~\mu} 
  + J_{\mu\rho} \bigl[ \cJ_{\al\be} \bigr]^\rho_{~\nu} 
\eql{commutators:J}  
\eeq
with $Z_\mu= P_\mu, K_\mu$.
The commutators~\eq{scaling:PK} and~\eq{commutators:J} with $[P_\mu, P_\nu]=0$ then imply that 
$[P_\mu, K_\nu] \propto J_{\mu\nu} + \eta_{\mu\nu} D$. 
We fix our sign and normalization conventions for $K_\mu$ as
\beq
[P_\mu, K_\nu] = 2\I (J_{\mu\nu} + \eta_{\mu\nu} D)
\,.\eql{commutator:PK}
\eeq
We can now continue~\eq{KPK1} as
\beq
\text{\eq{KPK1}}
= \e^{\I x \cdot P -\frac12 [\I x \cdot P , \, \I X]} \, \e^{\I X}
= \e^{\I x' \!\cdot P} \e^{\I X}
\eeq
with
\begin{align}
X &= 2 x^\mu b^\nu (J_{\mu\nu} + \eta_{\mu\nu} D)
\,,\eql{def:X}\\
x'^\mu &= x^\mu - x^2 b^\mu + 2(b \dt x) x^\mu
\,,\eql{action:K:x}
\end{align}
so \eq{KPK0} becomes
\beq
g \, \Omega(x,\xi) \, h^{-1}
= \e^{\I x' \!\cdot P} \, \e^{\I X} \e^{\I (\xi+b) \cdot K} \, h^{-1}
\,.
\eeq
This will give $\Omega(x', \xi')$ in~\eq{trans:Omega} if we choose $h = \e^{\I X}$ and identify $\xi'$ through
\beq
 h \, \e^{\I (\xi+b) \cdot K} h^{-1} = \e^{\I \xi' \cdot K}
\,.\eql{xi'}
\eeq
Unpacking $X$ in $h = \e^{\I X}$ by using~\eq{def:X}, we get
\beq
& h = \e^{\I \omega^{\al\be} J_{\al\be} / 2} \, \e^{\I \la D}
\,,\\
& \omega^{\al\be} = 2(x^\al b^\be - x^\be b^\al)
\,,\quad
\la = 2 b \dt x
\,.\eql{action:K:h}
\eeq
The action of $h$ is straightforward because $H$ is linearly realized.
For $g = h = \e^{\I\la D}$ in~\eq{trans:Omega}, \eq{scaling:PK} implies 
\beq
x'^\mu = \e^\la x^\mu
\,,\quad
\xi'^\mu = \e^{-\la} \xi^\mu
\eql{action:D}
\eeq
while for $g = h = \e^{\I\omega^{\al\be} \! J_{\al\be}/2}$ in~\eq{trans:Omega}, 
\eq{commutators:J} implies
\beq
x^{\prime\mu} = \La^\mu_{~\nu} x^\nu
\,,\quad
\xi^{\prime\mu} = \La^\mu_{~\nu} \xi^\nu
\,,\quad
\La = \e^{\I\omega^{\al\be} \! \cJ_{\al\be}/2\,}
\,.\eql{action:J}
\eeq
By applying~\eq{action:D} and~\eq{action:J} to~\eq{xi'},
we finally get
\beq
\xi'^\mu
= \xi^\mu + b^\mu + 2\I x^\al b^\be \bigl( [\cJ_{\al\be}]^\mu_{~\nu} + \I \eta_{\al\be} \de^\mu_\nu \bigr) \xi^\nu 
\,.\eql{action:K:xi}
\eeq
Thus, the action of $\dsp{\e^{\I b \cdot K}}$ is nonlinearly realized due to this shift of $\xi^\mu$ by $b^\mu$.
 
We now proceed to the field theory of broken conformal invariance with unbroken scale invariance.
The first step is to introduce NG modes for the broken generators $K_\mu$ by promoting the coset space coordinates $\xi^\mu$ to a field $\Xi^\mu(x)$.
The transformation~\eq{action:K:xi} is thus promoted to
\beq
\Xi'^\mu(x')
&= \Xi^\mu(x) + b^\mu\! 
+ 2\I x^\al b^\be \bigl( [\cJ_{\al\be}]^\mu_{~\nu} + \I \eta_{\al\be} \de^\mu_\nu \bigr) \Xi^\nu(x)
\!\!
\eql{trans:Xi}
\eeq
with $x'$ being related to $x$ as in~\eq{action:K:x}.
The scale transformation of $\xi^\mu$ in~\eq{action:D} becomes $\Xi'^\mu(x') = \e^{-\la} \Xi^\mu(x)$. This means that $\Xi^\mu$ has scaling dimension~$1$, because
\eq{action:D} tells us that the derivative scales as $\del'_\mu = \e^{-\la} \del_\mu$. 

Although we do not have to do so, we could express $\Xi^\mu(x)$ via a scalar field $\omega(x)$ as
\beq
\Xi_\mu(x) = \del_\mu \omega(x)
\,,\eql{omega}
\eeq
which exactly reproduces~\eq{trans:Xi} if we let $\omega$ transforms as
\beq
\omega'(x') = \omega(x) + b \dt x
\,.\eql{trans:omega}
\eeq
Since both $\Xi_\mu$ and $\del_\mu$ have scaling dimension~$1$, \eq{omega} tells us that $\omega$ has scaling dimension~$0$.

Unlike what we normally call the dilaton, however, $\omega$ is not a Nambu-Goldstone mode for dilation 
because scale invariance is unbroken.
In particular, $\omega$ never appears without a derivative acting on it but always comes in the combination $\Xi_\mu = \del_\mu \omega$.
Thus, unlike the normal dilaton, we cannot multiply a power of $\e^{\omega}$ to an operator to ``absorb'' its scaling dimension.
This, of course, is a manifestation of our assumption that scale invariance is unbroken and linearly realized. 

For any non-NG local operator $\mathcal{O}(x)$, 
the action of $\e^{\I b \cdot K}$ transforms $\mathcal{O}(x)$ to $\mathcal{O}'(x')$ as
\beq
\mathcal{O}'(x') = h \mathcal{O}(x)
\,,
\eeq
where $x'$ and $h$ are respectively given in~\eq{action:K:x} and~\eq{action:K:h}.
Substituting~\eq{action:K:h} explicitly, we obtain
\beq
\mathcal{O}'(x') 
= \mathcal{O}(x) 
+ 2\I x^\al b^\be (J_{\al\be} + \eta_{\al\be} D) \mathcal{O}(x)
\,.\eql{Phi'}
\eeq
Now, since $[D, J_{\mu\nu}] = 0$ and both $D$ and $J_{\al\be}$ are linearly realized,
we take $\mathcal{O}$ to be simultaneously an eigenstate of $D$ and in an irreducible Lorentz representation.
Then, denoting $\mathcal{O}$'s scaling dimension by~$\De$ and recalling $\del'_\mu = \e^{-\la} \del_\mu$, 
we have $\mathcal{O}'(x') = \e^{\I \la D} \mathcal{O}(x) = \e^{-\la\De} \mathcal{O}(x)$.
We thus have $D\mathcal{O} = \I\De\mathcal{O}$ and, hence, 
\beq
\mathcal{O}'(x') 
= \mathcal{O}(x)
+ 2\I x^\al b^\be (j_{\al\be} + \I\eta_{\al\be} \De) \mathcal{O}(x)
\,,\eql{trans:Phi}
\eeq
where $j_{\mu\nu}$ are the Lorentz generators for the representation that $\mathcal{O}$ is in.
Comparing \eq{trans:Xi} and~\eq{trans:omega} with~\eq{trans:Phi}, 
we again see that $\Xi^\mu$ and $\omega$ have $\De =1$ and~$0$, respectively.

We should stress that \eq{trans:Phi} is general and applies to any local non-NG operator.
So, if an operator $\mathcal{O}$ has scaling dimension~$\De$ and is in representation~$r$, 
the operator $\del_\mu\mathcal{O}$ should transform according to the appropriate paraphrase of~\eq{trans:Phi} for dimension~$\De+1$ and representation 
$\dsp{\text{(vector)} \!\otimes\! r}$.
But it does not; \eq{trans:Phi} with~\eq{action:K:x} gives
\beq
\del'_\mu \mathcal{O}'(x')
&= \del_\mu \mathcal{O}(x)
+ 2\I x^\al b^\be \bigl( j_{\al\be} \de_\mu^\nu + [\cJ_{\al\be}]_\mu^{~\nu} 
\\ &\phantom{=} \hspace{8em}
+ \I\eta_{\al\be} \de_\mu^\nu (\De + 1) \bigr) \del_\nu \mathcal{O}(x)
\\ &\phantom{=} \hspace{3.7em}
+ 2\I b^\be (j_{\mu\be} + \I\eta_{\mu\be} \De) \mathcal{O}(x)
\,.\eql{ord_del}
\eeq
The first two lines above are exactly how any object with dimension $\De +1 $ in representation $\dsp{\text{(vector)} \!\otimes\! r}$, such as $\del_\mu\mathcal{O}$, should transform according to the paraphrase of~\eq{trans:Phi}.
The last line, however, contradicts with it.

This internal contradiction can be resolved by making a rule that, upon promoting a scale invariant theory to a conformal theory, we must replace every $\del_\mu$ by a covariant derivative $\cD_\mu$,
where $\cD_\mu$ is constructed in such a way that the last line in~\eq{ord_del} is cancelled.
The shift by $b^\mu$ in~\eq{trans:Xi} allows us to arrange
such cancellation by adding $\dsp{-2\I\Xi^\be (j_{\mu\be} + \I\eta_{\mu\be}\De) \mathcal{O}}$ to $\del_\mu \mathcal{O}$.
The covariant derivative is thus given by
\beq
\cD_\mu = \del_\mu - 2\I \Xi^{\nu\!}(x) (j_{\mu\nu} + \I\eta_{\mu\nu} \De)
\,,\eql{cov_dev} 
\eeq
where $j_{\mu\nu}$ and $\De$ are respectively the Lorentz generators and scaling dimension of the object that the derivative is acting on.
Therefore, by replacing every $\del_\mu$ by $\cD_\mu$, any operator in any scale invariant theory can be promoted to a conformally covariant operator.

It is interesting to compare~\eq{trans:Phi} with how $\mathcal{O}$ would transform under unbroken conformal invariance.
In that case, if $\mathcal{O}$ is a primary operator (i.e., if $K_\mu\mathcal{O}(0) = 0$), 
it would transform exactly as in~\eq{trans:Phi}, because the expressions of $x'$ and $h$ in~\eq{action:K:x} and~\eq{action:K:h} do not depend on $\xi$ and hence are identical to those under unbroken conformal invariance.
If $\mathcal{O}$ is instead a descendant, there would be extra terms beyond those in~\eq{trans:Phi}.
Therefore, 
in conformal theories promoted from scale invariant theories, 
all non-NG operators transform as primaries.

An important special case of~\eq{trans:Phi} is when $\mathcal{O}$ is the Lagrangian $\cL$ if it exists.
Unbroken scale invariance requires $\cL$ to have $\De = d$ in $d$ spacetime dimensions.
Then, from~\eq{trans:Phi}, 
$\cL$ transforms as $\cL'(x') = \cL(x) - \dsp{2d \, (b \dt x) \, \cL(x)}$,
while \eq{action:K:x} gives $\| \del x' / \del x \| = 1 + \dsp{2d \, (b \dt x)}$.
The action, $\int \dd^d x \, \cL(x)$, is thus conformally invariant.
This by itself does not mean that scale invariance implies conformal invariance.
For that, we must also show that $\cL$ does not ``contain'' the NG mode. We will return to this shortly.

If the conformal-to-scale breaking is spontaneous, 
the NG mode must be dynamical. 
However, this is not possible in unitary theories in $d \neq 2$.
First, let us regard $\Xi_\mu$ as a fundamental degree of freedom rather than $\omega$.
Since $\Xi^\mu$ has scaling dimension~1, the mass term, $\Xi_\mu \Xi^\mu$, is forbidden for $d \neq 2$ by scale invariance,
which is unbroken by assumption.
In Poincar\'e invariant quantum field theories, 
unitarity requires a massless vector field to be equipped with a gauge invariance.
But the covariant derivative~\eq{cov_dev} cannot be made covariant also under a gauge transformation $\Xi_\mu \to \Xi_\mu + \del_\mu\chi$. 
Therefore, $\Xi_\mu$ cannot be dynamical in unitary theories.

One should also recall that unitarity would require the scaling dimension of a vector operator to be $\geq d-1$,
which would directly contradict with the fact that $\Xi_\mu$ has $\De = 1$, unless $d=2$.
This bound only relies on the conformal algebra, 
which does not care whether it is linearly or non-linearly realized,
so it applies to our promoted theories as well.
The bound can be violated if the vector operator is a gauge field, but $\Xi_\mu$ cannot be a gauge field as we just argued above.
We thus again see that a spontaneous breakdown of conformal to scale invariance would require the theory to be non-unitary. 

The unitarity bound argument can also be applied to $\omega$ if one wishes to regard $\omega$ as a fundamental degree of freedom.
Here, unitarity would imply that the scaling dimension of a scalar operator must be $\geq (d-2)/2$,
which would contradict with the fact that $\omega$ has $\De = 0$, unless $d=2$.
So, again, unitarity excludes the possibility of spontaneously breaking conformal to scale invariance.

For $d=2$, the unitary bound for $\omega$ is saturated. 
So, if the theory is unitary, $\omega$ is a free field and decouples from the theory.
Thus, promoting a scale invariant unitary theory in $d=2$ to a conformal theory actually does not change the theory at all. This is a simple demonstration of the fact that 
unbroken scale invariance implies unbroken conformal invariance for unitary theories in $d=2$.

We thus conclude that conformal invariance cannot be spontaneously broken to scale invariance in unitary theories.
If conformal symmetry is spontaneously broken, it must be broken all the way down to Poincar\'e.
Unitarity is essential in the argument above. 
This conclusion is also consistent with the observation of~\cite{Dymarsky:2015jia} that any scale but not conformally invariant sub-sector of a unitary conformal field theory is a free theory.

Whether the theory is unitary or not,
giving up $\omega$ and regarding $\Xi_\mu$ as fundamental allows ``non-minimal'' promotion of a scale invariant theory to a conformal theory beyond just $\del \to \cD$ by including operators made purely of $\Xi_\mu$, in a way that $\omega$ cannot.
To find such an operator, we look at the transformation of the derivative of $\Xi_\mu$:
\beq
\del'_\mu \Xi_\nu(x')
= \text{(cov.)} + 2( \eta_{\mu\nu} b \dt \Xi - b_\mu \Xi_\nu - b_\nu \Xi_\mu )
\,,\eql{delmuXi}
\eeq
where ``(cov.)'' denotes the terms transforming covariantly as expected from~\eq{trans:Phi}.
Since the non-covariant terms are symmetric in $\mu \leftrightarrow \nu$,
one way to make it covariant is given by
\beq
\cF_{\mu\nu} = \del_\mu \Xi_\nu - \del_\nu \Xi_\mu
\,.
\eeq
Another way is to notice that the covariant derivative for $\Xi$ can be formed as
\beq
\cD_\mu \Xi_\nu = \del_\mu \Xi_\nu - \eta_{\mu\nu} \Xi \dt \Xi + \Xi_\mu \Xi_\nu
\,.
\eeq
Since $\cF_{\mu\nu}$ has $\De = 2$, an example of non-minimal promotion would be to add $\cF_{\mu\nu} \cO^{\mu\nu}$ to the Lagrangian in $d$ dimensions if the theory has an operator $\cO^{\mu\nu}$ with $\dsp{\De = d-2}$. 
But if $\Xi_\mu$ is not fundamental and given as $\Xi_\mu = \del_\mu \omega$, 
$\cF_{\mu\nu}$ vanishes identically so such non-minimal promotion would not be possible.
If we give up unitarity and regard $\Xi_\mu$ as dynamical, 
the operator $\cF_{\mu\nu} \cF^{\mu\nu}$ has $\De = 4$ so it can be the kinetic term for $\Xi_\mu$ in $d = 4$. 

Non-minimal promotion is particularly interesting if we compare the present approach to those used in e.g.~\cite{Komargodski:2011vj, Luty:2012ww}, where a given field theory is formally rendered Weyl invariant by coupling it to a curved metric $g_{\mu\nu}(x) = \e^{-2\tau(x)} \eta_{\mu\nu}$.
This is classically equivalent to the present formalism with $\tau = 2\omega$ and appropriate field redefinitions.
At the quantum level, gravitational anomalies need to be taken into account to establish a mapping between these two approaches, which is beyond the scope of this work.
Here, we just note that non-minimal promotion via $\cF_{\mu\nu}$ does not seem to correspond to a simple operation, if any, in the $\tau$ language even classically.

If there exists a Lagrangian for a scale invariant theory that is also conformally invariant, 
the linear coupling of $\Xi^\mu$ in the promoted Lagrangian should decouple in the zero momentum limit of $\Xi^\mu$, 
that is, it should appear as $\Xi^\mu \del_\mu(\cdots)$.
(There is a subtle nuance to this statement in gauge theories to be discussed shortly).
This is because the action of $\e^{\I b \cdot K}$ in the original theory can be reproduced in the promoted theory by expanding to first order in $\Xi^\mu$ and set $\Xi^\mu = -b^\mu$, as is clear from the discussion leading to~\eq{cov_dev}.
Then, if the linear coupling of $\Xi^\mu$ in the promoted Lagrangian decouples in the zero momentum limit,
the action of $\e^{\I b \cdot K}$ on the original Lagrangian is a total derivative and hence is a symmetry.  

For example, consider a free massless scalar theory in $d$ spacetime dimensions:
\beq
\cL_0 = - (\del_\mu \phi^*) (\del^\mu \phi)
\,.
\eeq
If we promote $\cL_0$ by replacing $\del_\mu$ by $\cD_\mu$, we get
\beq
\cL 
&= - (\cD_\mu \phi^*) (\cD^\mu \phi)
\\
&= \cL_0 - (d-2) \Xi^\mu (\phi^* \del_\mu \phi + \phi \del_\mu \phi^*) + \cO(\Xi^2)
\,,
\eeq
where $\cD_\mu\phi = \del_\mu\phi + (d-2) \Xi_\mu\phi$ from~\eq{cov_dev}.
The linear coupling of $\Xi^\mu$ is to a total derivative $\del_\mu(\phi^*\phi)$,
indicating that the original theory, $\cL_0$, is conformally invariant.
For a free massless spin-$1/2$ fermion, we have $\bar{\psi} \bar{\sgm}^\mu \cD_\mu \psi = \bar{\psi} \bar{\sgm}^\mu \del_\mu \psi$ for all $d$ and $\Xi_\mu$ decouples completely.

We should note that $\del^\mu(\phi^*\phi)$ is the virial current~\cite{Coleman:1970je} of the free scalar theory.
It seems reasonable to conjecture that $\Xi_\mu$ couples to the virial current in scale invariant theories.
For \emph{conformally} invariant theories with Lagrangians, \cite{Kharuk:2017jwe} shows this is indeed the case classically.
A subtlety is that in the usual language of the stress-energy tensor it is often necessary to ``improve'' the tensor~\cite{Callan:1970ze, Coleman:1970je, Polchinski:1987dy}. 
It would be interesting to understand improvements in general without referring to the Lagrangian and at the quantum level.
(For the scalar example above, the improvement is to add a term $\propto \phi^2 D_\mu \Xi^\mu$ to cancel the $\cO(\Xi^2)$ term~\cite{Kharuk:2017jwe}.)

Another example of a scale invariant theory is given by a free U(1) gauge theory.
With gauge-fixing and ghost terms, the Lagrangian reads
\beq
\cL_0 = -\frac14 F_{\mu\nu} F^{\mu\nu} + \fr{B^2}{2\al} + B \del_\mu A^\mu -\bar{c}\, \del^2 c
\,,\eql{freeMaxwell}
\eeq
where $\al$ is a gauge-fixing parameter.
Since the scaling dimensions of $A_\mu$ is $(d-2)/2$,
we have
\beq
\cD_\mu A_\nu = \del_\mu A_\nu - 2\eta_{\mu\nu} \Xi \dt A + 2 A_\mu \Xi_\nu + (d-2) \Xi_\mu A_\nu
\,, 
\eeq
so promoting $F_{\mu\nu}$ and $B \del_\mu A^\mu$ respectively gives us
\begin{align}
\cD_\mu A_\nu - \cD_\nu A_\mu 
&= F_{\mu\nu} + (d - 4) (\Xi_\mu A_\nu - \Xi_\nu A_\mu)
\,,\eql{field_strength}
\\
B \cD_\mu A^\mu 
&= B \del_\mu A^\mu - d B \Xi_\mu A^\mu
\,.\eql{BDA}
\end{align}
For $c$ and $\bar{c}$, the free scalar analysis above makes it clear that the linear coupling of $\Xi^\mu$ from promoting $\bar{c}\, \del^2 c$ is a total derivative if and only if we assign an equal scaling dimension for $c$ and $\bar{c}$.
We see from \eq{field_strength} that the linear coupling of $\Xi^\mu$ from promoting $F_{\mu\nu} F^{\mu\nu}$ is not a total derivative in $d \neq 4$, while it completely vanishes in $d=4$.
Moreover, the promoted gauge kinetic term is not BRST invariant unless $d=4$.
In~\eq{BDA}, $\Xi_\mu$  couples to $BA^\mu$, which is not a total derivative for any $d$.
Only for $d=4$, however, the BRST invariance implies that the states created by $BA^\mu$ have a vanishing inner product with all physical states, so $\Xi_\mu$ does actually decouple.

Therefore, as is well known (e.g.,~\cite{Jackiw:2011vz, ElShowk:2011gz}), 
a free Maxwell theory in $d=4$ is conformally invariant,
while it is not in $d \neq 4$ (although there is a caveat that we will mention shortly). 
Moreover, in $d \neq 4$, the promoted theory loses BRST invariance and hence is not unitary.
The discussion above also makes it clear that, even in $d=4$, a free massless vector field without gauge invariance (i.e., without the ghost term in~\eq{freeMaxwell}) is not conformal, as originally observed in~\cite{Coleman:1970je}.
 
Finally, we should be careful as we may be ``fooled'' by some theories. 
One example is the Riva-Cardy model at the Weyl invariant (hence conformal) point~\cite{Riva:2005gd}.
This theory is given in Euclidean $d=2$ by
\beq
\cL_0 =  u_{\mu\nu} u^{\mu\nu} + \fr{k}{2} (\del_\mu A^\mu)^2 
\,,
\eeq
where $u_{\mu\nu} \equiv (\del_\mu A_\nu + \del_\nu A_\mu)/2$.
This theory is not unitary as it is not gauge invariant due to the lack of the ghost term.
When this Lagrangian is promoted,
$\Xi^\mu$ couples to $(k+2)(\del_\nu A^\nu) A_\mu -  A^\nu (\del_\nu A_\mu)$. 
This is not a total derivative except at $k=-3$ at which the theory is conformally invariant
(but not Weyl invariant, which does not contradict with~\cite{Polchinski:1987dy, Farnsworth:2017tbz} as the theory is not unitary). 
Interestingly, the theory is also conformal (and Weyl invariant) at $k=-1$.
This is because $\cL_0$ with $k=-1$ is just $(\del_\mu A_\nu) (\del^{\mu\!} A^\nu) / 2$ 
so $A_1$ and $A_2$ are actually two independent free scalars.
We missed this because we assumed that $A_\mu$ was a vector when specifying the $j_{\mu\nu}$ inside the $\cD_\mu$.
Another tricky theory is the free U(1) gauge theory in $d=3$, which is not conformal according to our diagnosis above. But it is dual to a free massless scalar, which is conformal, provided the theory is changed in a subtle way~\cite{ElShowk:2011gz} (so our diagnosis is still valid in a subtle way).
   
Needless to say, the challenge is to demonstrate the decoupling of $\Xi_\mu$ in interacting scale invariant theories without relying on the Lagrangians, but only assuming the four general properties that are violated by known examples of non-conformal scale invariant theories: unitarity, Poincar\'e invariance, discrete scaling dimensions, and the existence of scale current~\cite{Nakayama:2013is}.
A possible approach might be to apply the present formalism to operator product expansions (OPEs), 
in a similar way to the OPE study~\cite{Karananas:2017zrg} of nonlinearly realized conformal invariance broken to Poincar\'e.
It is possible that the peculiar property of promoted theories having only primaries and no descendants might play an important role.

\begin{acknowledgments}
The author thanks Markus Luty for numerous wonderful and insightful discussions about scale versus conformal invariance as well as for comments on the manuscript.
The author is supported by the US Department of Energy 
under grant DE-SC0010102.
\end{acknowledgments}


%


\begin{thebibliography}{26}%
\makeatletter
\providecommand \@ifxundefined [1]{%
 \@ifx{#1\undefined}
}%
\providecommand \@ifnum [1]{%
 \ifnum #1\expandafter \@firstoftwo
 \else \expandafter \@secondoftwo
 \fi
}%
\providecommand \@ifx [1]{%
 \ifx #1\expandafter \@firstoftwo
 \else \expandafter \@secondoftwo
 \fi
}%
\providecommand \natexlab [1]{#1}%
\providecommand \enquote  [1]{``#1''}%
\providecommand \bibnamefont  [1]{#1}%
\providecommand \bibfnamefont [1]{#1}%
\providecommand \citenamefont [1]{#1}%
\providecommand \href@noop [0]{\@secondoftwo}%
\providecommand \href [0]{\begingroup \@sanitize@url \@href}%
\providecommand \@href[1]{\@@startlink{#1}\@@href}%
\providecommand \@@href[1]{\endgroup#1\@@endlink}%
\providecommand \@sanitize@url [0]{\catcode `\\12\catcode `\$12\catcode
  `\&12\catcode `\#12\catcode `\^12\catcode `\_12\catcode `\%12\relax}%
\providecommand \@@startlink[1]{}%
\providecommand \@@endlink[0]{}%
\providecommand \url  [0]{\begingroup\@sanitize@url \@url }%
\providecommand \@url [1]{\endgroup\@href {#1}{\urlprefix }}%
\providecommand \urlprefix  [0]{URL }%
\providecommand \Eprint [0]{\href }%
\providecommand \doibase [0]{http://dx.doi.org/}%
\providecommand \selectlanguage [0]{\@gobble}%
\providecommand \bibinfo  [0]{\@secondoftwo}%
\providecommand \bibfield  [0]{\@secondoftwo}%
\providecommand \translation [1]{[#1]}%
\providecommand \BibitemOpen [0]{}%
\providecommand \bibitemStop [0]{}%
\providecommand \bibitemNoStop [0]{.\EOS\space}%
\providecommand \EOS [0]{\spacefactor3000\relax}%
\providecommand \BibitemShut  [1]{\csname bibitem#1\endcsname}%
\let\auto@bib@innerbib\@empty
\bibitem [{\citenamefont {Polchinski}(1988)}]{Polchinski:1987dy}%
  \BibitemOpen
  \bibfield  {author} {\bibinfo {author} {\bibfnamefont {J.}~\bibnamefont
  {Polchinski}},\ }\href {\doibase 10.1016/0550-3213(88)90179-4} {\bibfield
  {journal} {\bibinfo  {journal} {Nucl. Phys.}\ }\textbf {\bibinfo {volume}
  {B303}},\ \bibinfo {pages} {226} (\bibinfo {year} {1988})}\BibitemShut
  {NoStop}%
\bibitem [{\citenamefont {Zamolodchikov}(1986)}]{Zamolodchikov:1986gt}%
  \BibitemOpen
  \bibfield  {author} {\bibinfo {author} {\bibfnamefont {A.~B.}\ \bibnamefont
  {Zamolodchikov}},\ }\href@noop {} {\bibfield  {journal} {\bibinfo  {journal}
  {JETP Lett.}\ }\textbf {\bibinfo {volume} {43}},\ \bibinfo {pages} {730}
  (\bibinfo {year} {1986})},\ \bibinfo {note} {[Pisma Zh. Eksp. Teor.
  Fiz.43,565(1986)]}\BibitemShut {NoStop}%
\bibitem [{\citenamefont {Luty}\ \emph {et~al.}(2013)\citenamefont {Luty},
  \citenamefont {Polchinski},\ and\ \citenamefont {Rattazzi}}]{Luty:2012ww}%
  \BibitemOpen
  \bibfield  {author} {\bibinfo {author} {\bibfnamefont {M.~A.}\ \bibnamefont
  {Luty}}, \bibinfo {author} {\bibfnamefont {J.}~\bibnamefont {Polchinski}}, \
  and\ \bibinfo {author} {\bibfnamefont {R.}~\bibnamefont {Rattazzi}},\ }\href
  {\doibase 10.1007/JHEP01(2013)152} {\bibfield  {journal} {\bibinfo  {journal}
  {JHEP}\ }\textbf {\bibinfo {volume} {01}},\ \bibinfo {pages} {152} (\bibinfo
  {year} {2013})},\ \Eprint {http://arxiv.org/abs/1204.5221} {arXiv:1204.5221
  [hep-th]} \BibitemShut {NoStop}%
\bibitem [{\citenamefont {Dymarsky}\ \emph {et~al.}(2015)\citenamefont
  {Dymarsky}, \citenamefont {Komargodski}, \citenamefont {Schwimmer},\ and\
  \citenamefont {Theisen}}]{Dymarsky:2013pqa}%
  \BibitemOpen
  \bibfield  {author} {\bibinfo {author} {\bibfnamefont {A.}~\bibnamefont
  {Dymarsky}}, \bibinfo {author} {\bibfnamefont {Z.}~\bibnamefont
  {Komargodski}}, \bibinfo {author} {\bibfnamefont {A.}~\bibnamefont
  {Schwimmer}}, \ and\ \bibinfo {author} {\bibfnamefont {S.}~\bibnamefont
  {Theisen}},\ }\href {\doibase 10.1007/JHEP10(2015)171} {\bibfield  {journal}
  {\bibinfo  {journal} {JHEP}\ }\textbf {\bibinfo {volume} {10}},\ \bibinfo
  {pages} {171} (\bibinfo {year} {2015})},\ \Eprint
  {http://arxiv.org/abs/1309.2921} {arXiv:1309.2921 [hep-th]} \BibitemShut
  {NoStop}%
\bibitem [{\citenamefont {Dymarsky}\ \emph {et~al.}(2016)\citenamefont
  {Dymarsky}, \citenamefont {Farnsworth}, \citenamefont {Komargodski},
  \citenamefont {Luty},\ and\ \citenamefont {Prilepina}}]{Dymarsky:2014zja}%
  \BibitemOpen
  \bibfield  {author} {\bibinfo {author} {\bibfnamefont {A.}~\bibnamefont
  {Dymarsky}}, \bibinfo {author} {\bibfnamefont {K.}~\bibnamefont
  {Farnsworth}}, \bibinfo {author} {\bibfnamefont {Z.}~\bibnamefont
  {Komargodski}}, \bibinfo {author} {\bibfnamefont {M.~A.}\ \bibnamefont
  {Luty}}, \ and\ \bibinfo {author} {\bibfnamefont {V.}~\bibnamefont
  {Prilepina}},\ }\href {\doibase 10.1007/JHEP02(2016)099} {\bibfield
  {journal} {\bibinfo  {journal} {JHEP}\ }\textbf {\bibinfo {volume} {02}},\
  \bibinfo {pages} {099} (\bibinfo {year} {2016})},\ \Eprint
  {http://arxiv.org/abs/1402.6322} {arXiv:1402.6322 [hep-th]} \BibitemShut
  {NoStop}%
\bibitem [{\citenamefont {Yonekura}(2014)}]{Yonekura:2014tha}%
  \BibitemOpen
  \bibfield  {author} {\bibinfo {author} {\bibfnamefont {K.}~\bibnamefont
  {Yonekura}},\ }\href@noop {} {\  (\bibinfo {year} {2014})},\ \Eprint
  {http://arxiv.org/abs/1403.4939} {arXiv:1403.4939 [hep-th]} \BibitemShut
  {NoStop}%
\bibitem [{\citenamefont {Naseh}(2016)}]{Naseh:2016maw}%
  \BibitemOpen
  \bibfield  {author} {\bibinfo {author} {\bibfnamefont {A.}~\bibnamefont
  {Naseh}},\ }\href {\doibase 10.1103/PhysRevD.94.125015} {\bibfield  {journal}
  {\bibinfo  {journal} {Phys. Rev.}\ }\textbf {\bibinfo {volume} {D94}},\
  \bibinfo {pages} {125015} (\bibinfo {year} {2016})},\ \Eprint
  {http://arxiv.org/abs/1607.07899} {arXiv:1607.07899 [hep-th]} \BibitemShut
  {NoStop}%
\bibitem [{\citenamefont {Nakayama}(2015)}]{Nakayama:2013is}%
  \BibitemOpen
  \bibfield  {author} {\bibinfo {author} {\bibfnamefont {Y.}~\bibnamefont
  {Nakayama}},\ }\href {\doibase 10.1016/j.physrep.2014.12.003} {\bibfield
  {journal} {\bibinfo  {journal} {Phys. Rept.}\ }\textbf {\bibinfo {volume}
  {569}},\ \bibinfo {pages} {1} (\bibinfo {year} {2015})},\ \Eprint
  {http://arxiv.org/abs/1302.0884} {arXiv:1302.0884 [hep-th]} \BibitemShut
  {NoStop}%
\bibitem [{\citenamefont {Coleman}\ \emph {et~al.}(1969)\citenamefont
  {Coleman}, \citenamefont {Wess},\ and\ \citenamefont
  {Zumino}}]{Coleman:1969sm}%
  \BibitemOpen
  \bibfield  {author} {\bibinfo {author} {\bibfnamefont {S.~R.}\ \bibnamefont
  {Coleman}}, \bibinfo {author} {\bibfnamefont {J.}~\bibnamefont {Wess}}, \
  and\ \bibinfo {author} {\bibfnamefont {B.}~\bibnamefont {Zumino}},\ }\href
  {\doibase 10.1103/PhysRev.177.2239} {\bibfield  {journal} {\bibinfo
  {journal} {Phys. Rev.}\ }\textbf {\bibinfo {volume} {177}},\ \bibinfo {pages}
  {2239} (\bibinfo {year} {1969})}\BibitemShut {NoStop}%
\bibitem [{\citenamefont {Callan}\ \emph {et~al.}(1969)\citenamefont {Callan},
  \citenamefont {Coleman}, \citenamefont {Wess},\ and\ \citenamefont
  {Zumino}}]{Callan:1969sn}%
  \BibitemOpen
  \bibfield  {author} {\bibinfo {author} {\bibfnamefont {C.~G.}\ \bibnamefont
  {Callan}, \bibfnamefont {Jr.}}, \bibinfo {author} {\bibfnamefont {S.~R.}\
  \bibnamefont {Coleman}}, \bibinfo {author} {\bibfnamefont {J.}~\bibnamefont
  {Wess}}, \ and\ \bibinfo {author} {\bibfnamefont {B.}~\bibnamefont
  {Zumino}},\ }\href {\doibase 10.1103/PhysRev.177.2247} {\bibfield  {journal}
  {\bibinfo  {journal} {Phys. Rev.}\ }\textbf {\bibinfo {volume} {177}},\
  \bibinfo {pages} {2247} (\bibinfo {year} {1969})}\BibitemShut {NoStop}%
\bibitem [{\citenamefont {Low}\ and\ \citenamefont
  {Manohar}(2002)}]{Low:2001bw}%
  \BibitemOpen
  \bibfield  {author} {\bibinfo {author} {\bibfnamefont {I.}~\bibnamefont
  {Low}}\ and\ \bibinfo {author} {\bibfnamefont {A.~V.}\ \bibnamefont
  {Manohar}},\ }\href {\doibase 10.1103/PhysRevLett.88.101602} {\bibfield
  {journal} {\bibinfo  {journal} {Phys. Rev. Lett.}\ }\textbf {\bibinfo
  {volume} {88}},\ \bibinfo {pages} {101602} (\bibinfo {year} {2002})},\
  \Eprint {http://arxiv.org/abs/hep-th/0110285} {arXiv:hep-th/0110285 [hep-th]}
  \BibitemShut {NoStop}%
\bibitem [{\citenamefont {Hinterbichler}\ and\ \citenamefont
  {Khoury}(2012)}]{Hinterbichler:2011qk}%
  \BibitemOpen
  \bibfield  {author} {\bibinfo {author} {\bibfnamefont {K.}~\bibnamefont
  {Hinterbichler}}\ and\ \bibinfo {author} {\bibfnamefont {J.}~\bibnamefont
  {Khoury}},\ }\href {\doibase 10.1088/1475-7516/2012/04/023} {\bibfield
  {journal} {\bibinfo  {journal} {JCAP}\ }\textbf {\bibinfo {volume} {1204}},\
  \bibinfo {pages} {023} (\bibinfo {year} {2012})},\ \Eprint
  {http://arxiv.org/abs/1106.1428} {arXiv:1106.1428 [hep-th]} \BibitemShut
  {NoStop}%
\bibitem [{\citenamefont {Hinterbichler}\ \emph {et~al.}(2012)\citenamefont
  {Hinterbichler}, \citenamefont {Joyce},\ and\ \citenamefont
  {Khoury}}]{Hinterbichler:2012mv}%
  \BibitemOpen
  \bibfield  {author} {\bibinfo {author} {\bibfnamefont {K.}~\bibnamefont
  {Hinterbichler}}, \bibinfo {author} {\bibfnamefont {A.}~\bibnamefont
  {Joyce}}, \ and\ \bibinfo {author} {\bibfnamefont {J.}~\bibnamefont
  {Khoury}},\ }\href {\doibase 10.1088/1475-7516/2012/06/043} {\bibfield
  {journal} {\bibinfo  {journal} {JCAP}\ }\textbf {\bibinfo {volume} {1206}},\
  \bibinfo {pages} {043} (\bibinfo {year} {2012})},\ \Eprint
  {http://arxiv.org/abs/1202.6056} {arXiv:1202.6056 [hep-th]} \BibitemShut
  {NoStop}%
\bibitem [{\citenamefont {Salam}\ and\ \citenamefont
  {Strathdee}(1969)}]{Salam:1970qk}%
  \BibitemOpen
  \bibfield  {author} {\bibinfo {author} {\bibfnamefont {A.}~\bibnamefont
  {Salam}}\ and\ \bibinfo {author} {\bibfnamefont {J.~A.}\ \bibnamefont
  {Strathdee}},\ }\href {\doibase 10.1103/PhysRev.184.1760} {\bibfield
  {journal} {\bibinfo  {journal} {Phys. Rev.}\ }\textbf {\bibinfo {volume}
  {184}},\ \bibinfo {pages} {1760} (\bibinfo {year} {1969})}\BibitemShut
  {NoStop}%
\bibitem [{\citenamefont {Ivanov}\ and\ \citenamefont
  {Niederle}(1982{\natexlab{a}})}]{Ivanov:1981wn}%
  \BibitemOpen
  \bibfield  {author} {\bibinfo {author} {\bibfnamefont {E.~A.}\ \bibnamefont
  {Ivanov}}\ and\ \bibinfo {author} {\bibfnamefont {J.}~\bibnamefont
  {Niederle}},\ }\href {\doibase 10.1103/PhysRevD.25.976} {\bibfield  {journal}
  {\bibinfo  {journal} {Phys. Rev.}\ }\textbf {\bibinfo {volume} {D25}},\
  \bibinfo {pages} {976} (\bibinfo {year} {1982}{\natexlab{a}})}\BibitemShut
  {NoStop}%
\bibitem [{\citenamefont {Ivanov}\ and\ \citenamefont
  {Niederle}(1982{\natexlab{b}})}]{Ivanov:1981wm}%
  \BibitemOpen
  \bibfield  {author} {\bibinfo {author} {\bibfnamefont {E.~A.}\ \bibnamefont
  {Ivanov}}\ and\ \bibinfo {author} {\bibfnamefont {J.}~\bibnamefont
  {Niederle}},\ }\href {\doibase 10.1103/PhysRevD.25.988} {\bibfield  {journal}
  {\bibinfo  {journal} {Phys. Rev.}\ }\textbf {\bibinfo {volume} {D25}},\
  \bibinfo {pages} {988} (\bibinfo {year} {1982}{\natexlab{b}})}\BibitemShut
  {NoStop}%
\bibitem [{\citenamefont {Kharuk}(2018)}]{Kharuk:2017jwe}%
  \BibitemOpen
  \bibfield  {author} {\bibinfo {author} {\bibfnamefont {I.}~\bibnamefont
  {Kharuk}},\ }\href {\doibase 10.1103/PhysRevD.98.025006} {\bibfield
  {journal} {\bibinfo  {journal} {Phys. Rev.}\ }\textbf {\bibinfo {volume}
  {D98}},\ \bibinfo {pages} {025006} (\bibinfo {year} {2018})},\ \Eprint
  {http://arxiv.org/abs/1711.03411} {arXiv:1711.03411 [hep-th]} \BibitemShut
  {NoStop}%
\bibitem [{\citenamefont {Dymarsky}\ and\ \citenamefont
  {Zhiboedov}(2015)}]{Dymarsky:2015jia}%
  \BibitemOpen
  \bibfield  {author} {\bibinfo {author} {\bibfnamefont {A.}~\bibnamefont
  {Dymarsky}}\ and\ \bibinfo {author} {\bibfnamefont {A.}~\bibnamefont
  {Zhiboedov}},\ }\href {\doibase 10.1088/1751-8113/48/41/41FT01} {\bibfield
  {journal} {\bibinfo  {journal} {J. Phys.}\ }\textbf {\bibinfo {volume}
  {A48}},\ \bibinfo {pages} {41FT01} (\bibinfo {year} {2015})},\ \Eprint
  {http://arxiv.org/abs/1505.01152} {arXiv:1505.01152 [hep-th]} \BibitemShut
  {NoStop}%
\bibitem [{\citenamefont {Komargodski}\ and\ \citenamefont
  {Schwimmer}(2011)}]{Komargodski:2011vj}%
  \BibitemOpen
  \bibfield  {author} {\bibinfo {author} {\bibfnamefont {Z.}~\bibnamefont
  {Komargodski}}\ and\ \bibinfo {author} {\bibfnamefont {A.}~\bibnamefont
  {Schwimmer}},\ }\href {\doibase 10.1007/JHEP12(2011)099} {\bibfield
  {journal} {\bibinfo  {journal} {JHEP}\ }\textbf {\bibinfo {volume} {12}},\
  \bibinfo {pages} {099} (\bibinfo {year} {2011})},\ \Eprint
  {http://arxiv.org/abs/1107.3987} {arXiv:1107.3987 [hep-th]} \BibitemShut
  {NoStop}%
\bibitem [{\citenamefont {Coleman}\ and\ \citenamefont
  {Jackiw}(1971)}]{Coleman:1970je}%
  \BibitemOpen
  \bibfield  {author} {\bibinfo {author} {\bibfnamefont {S.~R.}\ \bibnamefont
  {Coleman}}\ and\ \bibinfo {author} {\bibfnamefont {R.}~\bibnamefont
  {Jackiw}},\ }\href {\doibase 10.1016/0003-4916(71)90153-9} {\bibfield
  {journal} {\bibinfo  {journal} {Annals Phys.}\ }\textbf {\bibinfo {volume}
  {67}},\ \bibinfo {pages} {552} (\bibinfo {year} {1971})}\BibitemShut
  {NoStop}%
\bibitem [{\citenamefont {Callan}\ \emph {et~al.}(1970)\citenamefont {Callan},
  \citenamefont {Coleman},\ and\ \citenamefont {Jackiw}}]{Callan:1970ze}%
  \BibitemOpen
  \bibfield  {author} {\bibinfo {author} {\bibfnamefont {C.~G.}\ \bibnamefont
  {Callan}, \bibfnamefont {Jr.}}, \bibinfo {author} {\bibfnamefont {S.~R.}\
  \bibnamefont {Coleman}}, \ and\ \bibinfo {author} {\bibfnamefont
  {R.}~\bibnamefont {Jackiw}},\ }\href {\doibase 10.1016/0003-4916(70)90394-5}
  {\bibfield  {journal} {\bibinfo  {journal} {Annals Phys.}\ }\textbf {\bibinfo
  {volume} {59}},\ \bibinfo {pages} {42} (\bibinfo {year} {1970})}\BibitemShut
  {NoStop}%
\bibitem [{\citenamefont {Jackiw}\ and\ \citenamefont
  {Pi}(2011)}]{Jackiw:2011vz}%
  \BibitemOpen
  \bibfield  {author} {\bibinfo {author} {\bibfnamefont {R.}~\bibnamefont
  {Jackiw}}\ and\ \bibinfo {author} {\bibfnamefont {S.~Y.}\ \bibnamefont
  {Pi}},\ }\href {\doibase 10.1088/1751-8113/44/22/223001} {\bibfield
  {journal} {\bibinfo  {journal} {J. Phys.}\ }\textbf {\bibinfo {volume}
  {A44}},\ \bibinfo {pages} {223001} (\bibinfo {year} {2011})},\ \Eprint
  {http://arxiv.org/abs/1101.4886} {arXiv:1101.4886 [math-ph]} \BibitemShut
  {NoStop}%
\bibitem [{\citenamefont {El-Showk}\ \emph {et~al.}(2011)\citenamefont
  {El-Showk}, \citenamefont {Nakayama},\ and\ \citenamefont
  {Rychkov}}]{ElShowk:2011gz}%
  \BibitemOpen
  \bibfield  {author} {\bibinfo {author} {\bibfnamefont {S.}~\bibnamefont
  {El-Showk}}, \bibinfo {author} {\bibfnamefont {Y.}~\bibnamefont {Nakayama}},
  \ and\ \bibinfo {author} {\bibfnamefont {S.}~\bibnamefont {Rychkov}},\ }\href
  {\doibase 10.1016/j.nuclphysb.2011.03.008} {\bibfield  {journal} {\bibinfo
  {journal} {Nucl. Phys.}\ }\textbf {\bibinfo {volume} {B848}},\ \bibinfo
  {pages} {578} (\bibinfo {year} {2011})},\ \Eprint
  {http://arxiv.org/abs/1101.5385} {arXiv:1101.5385 [hep-th]} \BibitemShut
  {NoStop}%
\bibitem [{\citenamefont {Riva}\ and\ \citenamefont
  {Cardy}(2005)}]{Riva:2005gd}%
  \BibitemOpen
  \bibfield  {author} {\bibinfo {author} {\bibfnamefont {V.}~\bibnamefont
  {Riva}}\ and\ \bibinfo {author} {\bibfnamefont {J.~L.}\ \bibnamefont
  {Cardy}},\ }\href {\doibase 10.1016/j.physletb.2005.07.010} {\bibfield
  {journal} {\bibinfo  {journal} {Phys. Lett.}\ }\textbf {\bibinfo {volume}
  {B622}},\ \bibinfo {pages} {339} (\bibinfo {year} {2005})},\ \Eprint
  {http://arxiv.org/abs/hep-th/0504197} {arXiv:hep-th/0504197 [hep-th]}
  \BibitemShut {NoStop}%
\bibitem [{\citenamefont {Farnsworth}\ \emph {et~al.}(2017)\citenamefont
  {Farnsworth}, \citenamefont {Luty},\ and\ \citenamefont
  {Prilepina}}]{Farnsworth:2017tbz}%
  \BibitemOpen
  \bibfield  {author} {\bibinfo {author} {\bibfnamefont {K.}~\bibnamefont
  {Farnsworth}}, \bibinfo {author} {\bibfnamefont {M.~A.}\ \bibnamefont
  {Luty}}, \ and\ \bibinfo {author} {\bibfnamefont {V.}~\bibnamefont
  {Prilepina}},\ }\href {\doibase 10.1007/JHEP10(2017)170} {\bibfield
  {journal} {\bibinfo  {journal} {JHEP}\ }\textbf {\bibinfo {volume} {10}},\
  \bibinfo {pages} {170} (\bibinfo {year} {2017})},\ \Eprint
  {http://arxiv.org/abs/1702.07079} {arXiv:1702.07079 [hep-th]} \BibitemShut
  {NoStop}%
\bibitem [{\citenamefont {Karananas}\ and\ \citenamefont
  {Shaposhnikov}(2018)}]{Karananas:2017zrg}%
  \BibitemOpen
  \bibfield  {author} {\bibinfo {author} {\bibfnamefont {G.~K.}\ \bibnamefont
  {Karananas}}\ and\ \bibinfo {author} {\bibfnamefont {M.}~\bibnamefont
  {Shaposhnikov}},\ }\href {\doibase 10.1103/PhysRevD.97.045009} {\bibfield
  {journal} {\bibinfo  {journal} {Phys. Rev.}\ }\textbf {\bibinfo {volume}
  {D97}},\ \bibinfo {pages} {045009} (\bibinfo {year} {2018})},\ \Eprint
  {http://arxiv.org/abs/1708.02220} {arXiv:1708.02220 [hep-th]} \BibitemShut
  {NoStop}%
\end{thebibliography}
\end{document}